\documentstyle[11pt,newpasp,twoside,epsf]{article}
\begin{document}
\title{The Galactic Bar}
 \author{Michael R. Merrifield}
\affil{University of Nottingham, Nottingham NG7 2RD, UK}

\begin{abstract}
Like the majority of spiral galaxies, the Milky Way contains a central
non-axisymmetric bar component.  Our position in the Galactic plane
renders it rather hard to see, but also allows us to make measurements
of the bar that are completely unobtainable for any other system.
This paper reviews the evidence for a bar that can be gleaned from the
many extensive surveys of both gas and stars in the Milky Way.  We
introduce some simplified models to show how the basic properties of
the bar can be inferred in a reasonably robust manner despite our
unfavorable location, and how the complex geometry can be used to our
advantage to obtain a unique three-dimensional view of the bar.  The
emerging picture of the Galactic bar is also placed in the broader
context of current attempts to understand how such structures form and
evolve in spiral galaxies.
\end{abstract}

\section{Introduction}
The recognition that some galaxies contain central rectangular bars
dates back at least as far as Curtis' (1918) classification of a
number of such systems as ``$\phi$-type spirals.''  This fundamental
property of disk galaxies was codified in the form we know today in
Hubble's (1926) classic ``tuning fork'' diagram, where barred and
unbarred spirals form the prongs of the fork.  Over the succeeding
century, observations have improved, and in particular most recently
infrared data have allowed us to view the stellar components of
galaxies almost free from the confusing effects of dust obscuration.
This mass of data has shown that bars are very common features, found
in over a half of all spiral galaxies (Whyte et al.\ 2002).

\begin{figure}
\plotfiddle{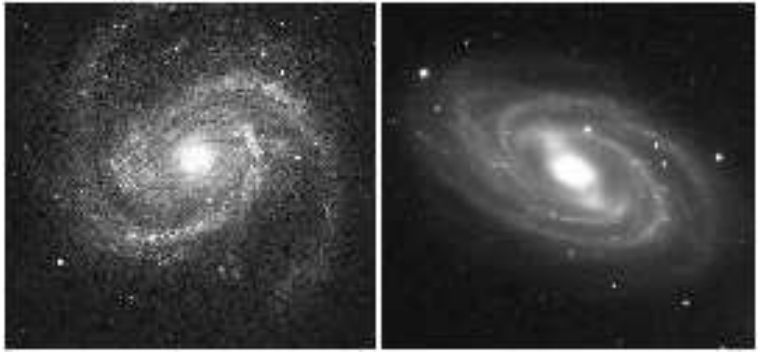}{2.5in}{0}{35}{35}{-150}{0}
\caption{Images of two nearby Sbc galaxies, the unbarred M100 and the
barred M109.  Which would the Milky Way resemble to an intergalactic
tourist? (Images from D. Malin and NOAO.)}
\end{figure}

It should therefore come as no particular surprise to find that the
Milky Way is also a barred system.  However, the commonness of such
galaxies also raises the question of why one should care whether the
Milky Way is barred or not.  There are, after all, plenty of other
galaxies that are oriented much more favorably than the Milky Way: not
only is our own galaxy edge-on, rendering any bar almost invisible,
but different parts of the Milky Way are different distances away,
greatly complicating the geometry.  Why not learn all there is to know
about bars by studying the plethora of more accessible external
systems?  In fact, there are several good reasons why we might want to
know about the Milky Way's bar.  First, it is natural to be
inquisitive about our own neighborhood, and want to know what our
galaxy would look like to an outsider (see Figure~1).  Second, we have
a mass of detailed information about the Milky Way that we have only
been able to obtain because of its proximity.  In order to put these
data about small-scale structure, star streams, stellar populations,
star formation, etc, into the broader context of galaxy evolution, we
need to quantify the global properties of the galaxy in which they
were found.  Finally, and perhaps most importantly, although our
position within the Galaxy complicates the understanding of its
structure, it also provides us with information on the
three-dimensional properties of its bar that is not available in any
other system.  Since we know very little about the three-dimensional
shapes of bars, the Milky Way offers a unique laboratory for studying
the structure of these systems.

In fact, our lack of understanding of their three-dimensional
structure even leads to some ambiguity as to what exactly is meant by
the term ``bar.'' In particular, it is not clear whether one should
draw a distinction between a spheroidal (possibly triaxial) central
bulge and a bar component.  The different nomenclature may just
reflect the limitations of our two-dimensional views of these systems:
a triaxial structure that would be identified as a bar in a face-on
system would be classified as a bulge component if the same system
were viewed edge on.  Indeed, there is now strong indirect evidence
that bulges and bars are at the very least intimately related (Bureau
\& Freeman 1999, Merrifield \& Kuijken 1999).  In order not to
prejudge the issue, this review will not draw any absolute distinction
between the various putative components of the Galaxy in its central
few kiloparsecs, but rather will look at the total distribution of
stars and gas to see how far they are from axisymmetry, and hence how
strongly the Milky Way is barred in that global sense.

In addition to referring the interested reader to the sophisticated
treatment of the wealth of data available for such studies, we will
attempt to develop some simple models that illustrate the particular
challenges of investigating the bar in our own galaxy and the unique
possibilities that its geometry admits.  Section~2 looks at the
properties of the bar that can be derived from the Milky Way's gaseous
components, while Section~3 explores the stellar distribution.  In
Section~4, we place the Milky Way in the context of other barred
galaxies, and in Section~5 we take a brief look at what the future may
hold for studies of the Galactic bar.

\begin{figure}
\plotfiddle{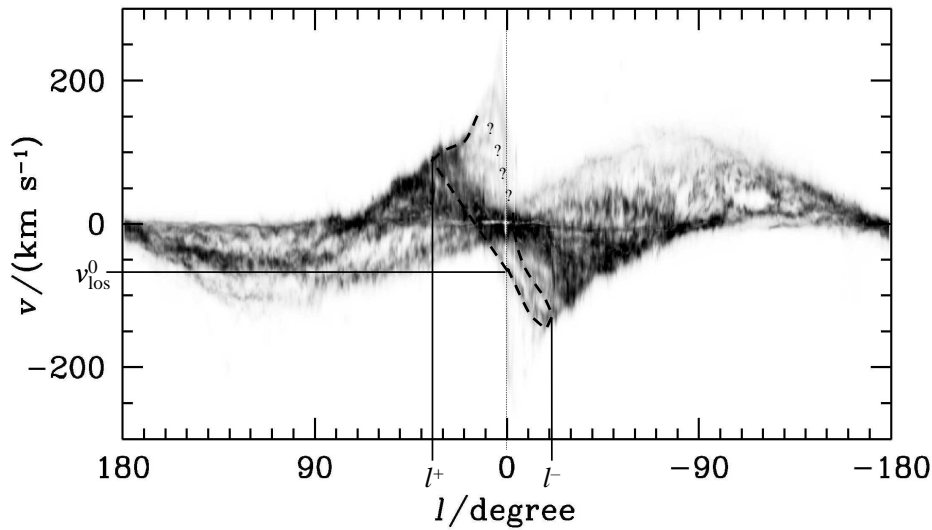}{2.8in}{0}{45}{45}{-200}{-30}
\caption{Plot showing the distribution of atomic hydrogen in the plane
of the Milky Way as a function of Galactic longitude and line-of-sight
velocity, as indicated by its 21cm emission.  The ``3 kpc arm'' is
highlighted, along with its important measurable parameters; the part
of the feature that is too faint to trace reliably is indicated by
question marks. (Adapted from Binney \& Merrifield 1998.)}
\end{figure}

\section{The Bar in Gas}
Although most people think about distortions in starlight when
considering bars, the earliest evidence that the Milky Way is barred
came from it gaseous component.  This is not surprising, as the Milky
Way is much like an external system viewed edge-on, and in such cases
a bar is almost impossible to detect photometrically.  The best clues
to the existence of a bar in an edge-on system come from the
kinematics of its gaseous components, where the non-axisymmetric bar
potential induces non-circular orbits in the gas.

In the case of the Milky Way, as Figure~2 shows, a plot of
line-of-sight velocity versus Galactic longitude (an ``$l$-$v$
diagram'') for the HI gas in the plane of the Galaxy reveals clear
signatures of a non-axisymmetric distribution.  There is a significant
asymmetry between the gas properties at positive and negative
longitudes, which is inconsistent with an axisymmetric disk.  Further
evidence for non-circular motions comes from the non-zero
line-of-sight velocities of some of the HI gas at zero longitude: if
the gas were following circular orbits, then all of its motion should
be transverse to the line of sight at this point.

The strongest feature that illustrates these properties is classically
known as the ``3~kpc expanding arm,'' because of its approximate
radial location in the Galaxy.  This feature, highlighted in Figure~2,
is asymmetric about the center of the Galaxy, lying between Galactic
longitudes of $l^- \sim -20^\circ$ and $l^+ \sim +35^\circ$.  Tracing
it through the center of the Galaxy reveals that it crosses zero
longitude at a velocity of $v_{\rm los}^0 = -53\, {\rm km}\,{\rm
s}^{-1}$; presumably this loop in the $l$--$v$ diagram re crosses $l=0$
at positive velocity, but this part of the feature is too faint to
trace reliably.  This non-zero line-of-sight velocity is inconsistent
with an axisymmetric picture of the Galaxy with gas on circular
orbits, as all gas motion in this direction should be transverse to
the line of sight.  It was this observation that led to the idea that
the feature might be an expanding arm of material flung out from the
center of the Galaxy in our direction.  As we will see below, however,
it is now understood as arising from the non-circular motions that
occur in bars.

The interpretation of this feature as a bar is usually attributed to
de Vaucouleurs (1964), who first made such a connection.  However, his
description of ``focalized vortices plunging toward the nucleus'' is
not one that would be recognized today.  It is therefore perhaps
unsurprising that the properties he inferred for the bar turn out to
be erroneous: even the side of the bar that lies closest to us was
incorrectly identified.  The earliest accurate modeling of this
feature seems to have been undertaken by Shane (1971)\footnote{This
paper is seldom noticed because it has the misfortune to be in one of
the few volumes that is missing from the ADS, so is passed over in
many literature reviews.}, who pointed out that this feature could be
explained by an elliptical ring of gas with its major axis oriented at
an angle of $\phi \sim 20^\circ$ to the line of sight, and a
short-to-long axis ratio of $\sim 0.6$.  Subsequent analysis (e.g.\
Peters 1975) confirmed this result, and even recent full gas-dynamical
simulations (e.g.\ Fux 1999) all find that a feature of this kind is
required to fit the observed $l$-$v$ diagram.

In order to obtain some insight into the reason for the robustness of
this result, it is instructive to see if it can be understood on the
basis of a simple calculation.  Specifically, can one go from the
three basic observed quantities, $l^+$, $l^-$ and $v_{\rm los}^0$ to
obtain the three parameters of the ellipse, $b$, $a$ and $\phi$?  The
values of the spatial coordinates, $l^+$ and $l^-$, clearly do not
provide enough information to solve for all three ellipse parameters,
but obviously provide some constraint: the fact that they are not
equal rules out an axisymmetric feature.  One can therefore narrow
down the range of possible ellipses using $l^+$ and $l^-$, and, for
example, solve for the axis ratio $b/a$ as a function of the feature's
angle $\phi$ to the line of sight.  This constraint is illustrated in
Figure~3.

\begin{figure}
\plotfiddle{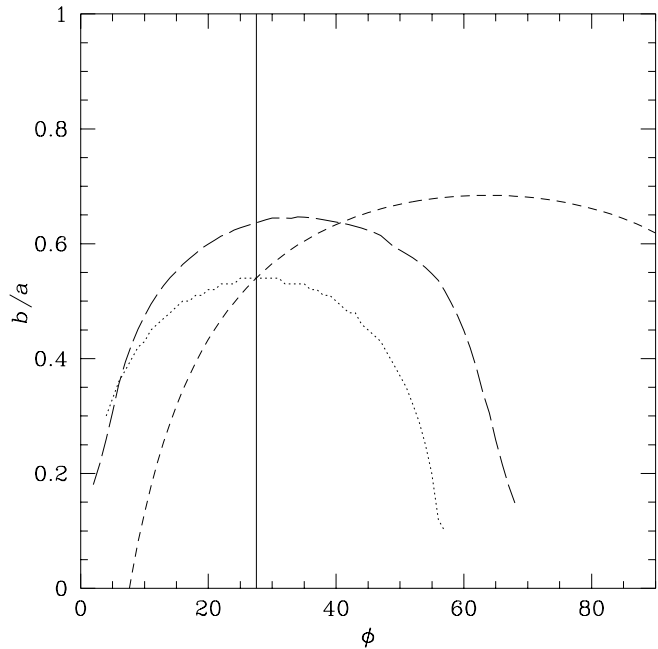}{3.2in}{0}{45}{45}{-140}{-70}
\caption{Constraints on in-plane flattening and position angle of the
Galactic bar relative to the Sun--Galactic-center line.  The dotted
line shows the constraint provided by the asymmetry in the extent of
the 3~kpc arm between positive and negative Galactic longitudes, while
the short-dashed line shows the simple kinematic constraint derived
from this feature.  The intersection of these lines implies that the
3~kpc gas ring is oriented at an angle of $\phi \sim 25^\circ$, with a
short-to-long axis ratio of $b/a \sim 0.5$; since the intersection
occurs at close to the maximum of the dotted line, the significant
uncertainties in the analysis producing the dashed line should have
little impact on the inferred value for $b/a$.  The long-dashed line
shows the constraint on the photometric bar provided by its asymmetry
about the Galactic center; again, the shape of this curve means that
the derived value of $b/a \sim 0.6$ for the stellar component is
robust against uncertainties in $\phi$.}
\end{figure}

A further constraint comes from the non-zero velocity observed at zero
longitude.  This observation can be interpreted rather crudely in the
epicyclic approximation, in which the elliptical orbit of the gas is
constructed from a parent circular orbit of angular frequency $\Omega$
and radius $R_{\rm bar}$ upon which an elliptical perturbation of
radial amplitude $X$ and angular frequency $\kappa$ is superimposed
(Binney \& Tremaine 1987).  At an arbitrary longitude, the
line-of-sight velocity of gas following such motions is an ugly
mixture of components of the circular motion plus the radial and
tangential perturbations.  However, at $l=0$ the only line-of-sight
component is the radial perturbation, which must produce the observed
value of $v_{\rm los}^0$.  Writing the radial coordinate of gas as $R
= R_{\rm bar} + X\cos\kappa t$, we find that $\dot R =
-X\kappa\sin\kappa t$.  The situation is further simplified if it is
assumed that the speed of circular orbits in the Milky Way, $v_c$, is
approximately constant, as observed in most external spiral galaxies
(although such an assumption must break down at small radii where the
rotation speed must drop toward zero).  For such a flat rotation
curve, $\kappa = \sqrt{2}\Omega = \sqrt{2} v_c/R_{\rm bar}$ (Binney \&
Tremaine 1987).  Putting all this information together, we find that
\begin{equation}
{X \over R_{\rm bar}} = {-v_{\rm los}^0 \over \sqrt{2}v_c} \times
                                           {1 \over \sin(\sqrt{2} \phi)},
\end{equation}
and hence the ellipse's axis ratio
\begin{equation}
{b \over a} = {R_{\rm bar} - X \over R_{\rm bar} + X} 
                            = {1 - X/R_{\rm bar} \over 1 + X/R_{\rm bar}},
\end{equation}
which is just a function of $\phi$.  This function is also plotted in
Figure~3.  

The spatial and the kinematic constraints are both met where the lines
intersect.  It is therefore apparent that a self-consistent model
requires an ellipse of gas with an axis ratio of $b/a \sim 0.55$ at an
angle to the line-of-sight of $\phi \sim 25^\circ$.  A number of
somewhat dubious approximations have gone into this argument -- the
epicyclic approximation has been pushed well beyond the perturbative
regime where it is strictly valid, for example.  However, full
hydrodynamical modeling of the Galactic $l$--$v$ diagrams in HI and CO
come up with essentially identical parameters for this gas ring (Fux
1999).  If this feature is interpreted as a signature of the Galactic
bar, then we have a measure of its angle to the line-of-sight; as we
shall see below, however, the axis ratio requires further
interpretation.

\section{The Bar in Stars}
Seeing the stellar bar in the Milky Way presents particular
difficulties.  As has already been mentioned, it is viewed from a most
unfavorable direction, but in addition it is seen through many
magnitudes of optical extinction, and it covers many square degrees on
the sky.  These latter challenges were finally overcome when the COBE
map of the sky was published, and Galactic astronomers began analyzing
the parts of the data that the cosmologists discarded (see Figure~4).
The COBE satellite's infrared coverage spanned the optimal point in
the spectrum where emission is still dominated by stars yet extinction
is minimized, and the all-sky coverage of the survey meant that the
large extent of the structure was no longer a problem, so finally some
photometric insight into the bar's properties could be obtained.

\begin{figure}
\plotfiddle{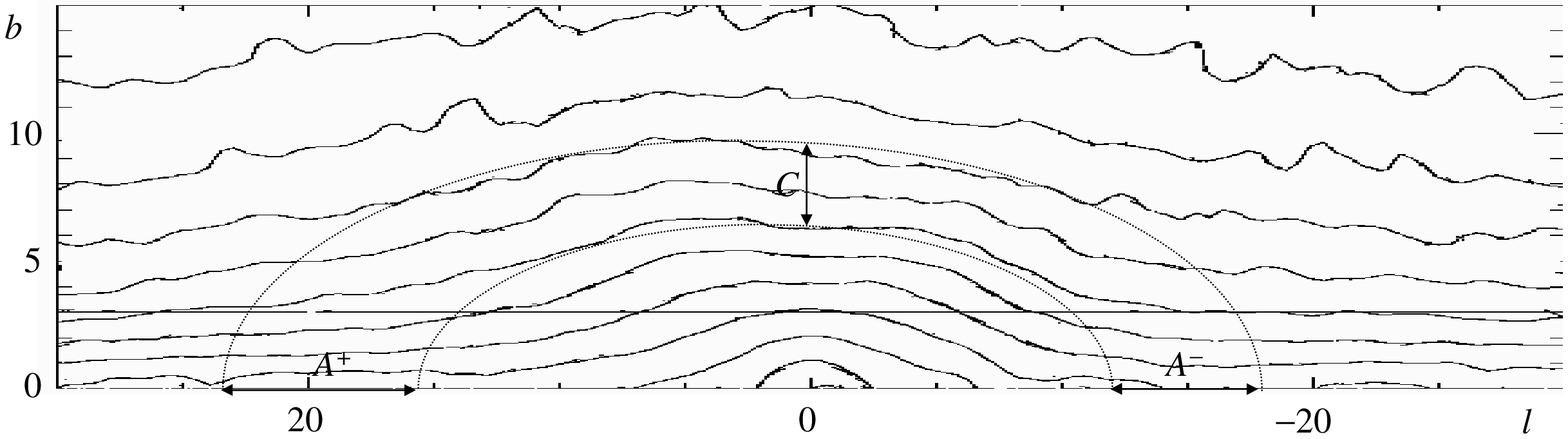}{1.45in}{0}{45}{45}{-200}{-130}
\caption{De-reddened infrared map of the center of the Galaxy as seen
by COBE.  Approximate ellipse fits to contours away from the Galactic
plane, separated by one $e$-folding in surface brightness, illustrate
the asymmetry caused by the Galactic bar.  The measured scalelengths
are indicated. (Adapted from Fux 1997.)}
\end{figure}

In external edge-on galaxies, bars are completely undetectable from
photometric data.  The only clues to their presence come from indirect
indicators such as a plateau in the light distribution at small radii
(e.g.\ de Carvalho \& da Costa 1987).  However, such features could
equally well be axisymmetric disk features, so no firm conclusions can
be drawn.  One can do rather better if one measures optical spectra
along the major axes of such galaxies, as both the stellar kinematics
(as determined from absorption-line spectra) and gas kinematics (as
seen in emission-line spectra) show complex structure in the
line-of-sight velocity distribution as a function of position (Kuijken
\& Merrifield 1995), directly analogous to the structure seen in
$l$--$v$ diagram of the Milky Way (see Figure~2).  It was through such
studies that a link was established between the presence of a bar and
boxy, or even double-lobed, isophotes in the central bulges of these
systems (Bureau \& Freeman 1999, Merrifield \& Kuijken 1999).  Such a
link was predicted by numerical simulations, which showed that
initially-thin bar structures should buckle perpendicular to their
plane, creating triaxial objects with a double-lobed shape when viewed
edge-on (e.g.\ Combes \& Sanders 1981).  The double-lobed structure
apparent in some of the isophotes shown in Figure~4 therefore provides
indirect evidence that the Milky Way is barred.

In the case of the Milky Way, however, we can obtain rather more
direct evidence for the presence of a bar.  Generally speaking, the
complex geometry involved in observing a galaxy from the inside
greatly hinders the interpretation of the data.  On this occasion,
though, it actually helps.  The finite distance to the bar means that
there is a parallax effect: the two ends of the bar are effectively
viewed from different directions, causing the projected scale-length
of the light distribution to be different on the two sides of the
Galactic center.  This asymmetry is clearly visible in the map of the
light distribution shown in Figure~4.

Perhaps unsurprisingly, there is not enough information in such a
two-dim\-ensional map to derive a unique model for the
three-dimensional distribution of starlight, even if quite strong
symmetries are imposed on the bar [as highlighted by Zhao (2000)].
Nonetheless, a fairly consistent view has emerged as to the basic
structural parameters of the bar using the observed asymmetry in the
photometry, and once again a simple model provides a useful
illustration as to why these parameters are fairly robustly measured.

The simplest plausible model for the bar is a triaxial Gaussian
ellipsoid, with principal axes of scalelengths $a$, $b$, and $c$ in
the $x$, $y$ and $z$ directions, respectively (where $z$ is the
distance perpendicular to the Galactic plane.  If viewed edge-on from
a large distance at an angle $\phi$ to the $x$-axis, the projected
distribution of light would also be Gaussian with a scale length $c$
perpendicular to the plane and a scale length
\begin{equation}
A(\phi) = \left[{{\sin^2\phi \over a^2} + {\cos^2\phi \over b^2} 
       - {\sin^2\phi\cos^2\phi\left({1 \over a^2} - {1 \over b^2}\right)
       \over {{\cos^2\phi \over a^2} + {\sin^2\phi \over b^2}}}}\right]^{1/2}
\end{equation}
in the projected plane.  However, since we are not an infinite
distance away, observations at $\pm l$ observe the bar from somewhat
different directions, which will yield different projected
scalelengths, $A^+ = A(\phi + l)$ and $A^- = A(\phi - l)$.  Thus, we
essentially have two observables at any given longitude,
$\overline{A}(l) = (A^+ + A^-)/2$ and $f(l) = A^+/A^-$; at $|l| \sim
15^\circ$, for example, we can see from Figure~4 that $\overline{A}
\sim 6.0^\circ$ and $f \sim 1.25$.  There are three in-plane physical
quantities that we would like to derive -- $a$, $b$ and $\phi$ -- so
clearly the problem is under-determined, illustrating the degeneracy in
deprojecting from two dimensions to three.  However, we can solve for
$a$ and $b/a$ as functions of $\phi$, as we did above for the gaseous
component; the resulting values for $b/a$ as a function of $\phi$ are
also shown in Figure~3.

If we adopt the angle of $\phi \sim 25^\circ$, as derived from the
gaseous component, we see that the stellar component must have an axis
ratio of $b/a \sim 0.6$.  Folding in the value of the vertical
scaleheight, $c$, as measured in Figure~4, we find that the bar is a
triaxial structure with axis ratios $a : b : c \sim 1 : 0.6 : 0.4$.
Again, although this result has been derived on the basis of a
simplified model, it reproduces the results of much more sophisticated
calculations: Fux's (1999) matching of N-body simulations to the COBE
map and Binney, Gerhard \& Spergel's (1997) direct deprojection of the
same data both predict aspect ratios of $a : b : c \sim 1 : 0.6 :
0.4$.  This good agreement must occur because the amount of
information on the structure of the bar is rather limited in the
two-dimensional COBE map, and all the various modeling techniques
latch onto essentially the same constraint arising from the asymmetry
in the distribution of starlight.  

Note that this analysis uses all the light from the central region,
and does not attempt to separate out distinct bar and bulge
components.  The fact that it is possible to produce a coherent
picture of the bar region without a separate axisymmetric bulge
component adds further weight to the argument that these features
should be viewed as different projections of the same physical object
rather than as distinct entities.  It is also worth noting that this
measurement illustrates why it is worth persisting with the
challenging task of studying the Milky Way's bar: such a fundamental
measurement of the three-dimensional shape of a bar has not been
possible in any other system.

One use that we can immediately make of this unique measurement is to
compare the shape of the bar in gas to that in the stellar component.
Since the gas disk is essentially confined to the Galactic plane, the
stellar bar is clearly rounder perpendicular to the plane than the
gaseous bar.  It is also apparent from Figure~3 that the stellar
component is also rounder in the plane, with an axis ratio of $b/a
\sim 0.6$ as compared to the value of $\sim 0.5$ for the gaseous
component.  This difference in shape can be straightforwardly
understood if the stars do not lie on the closed orbits that the gas
follows, but instead oscillate about these orbits, adding a more
random component to the mean streaming, thus puffing out the structure
of the stellar bar.

\begin{figure}
\plotfiddle{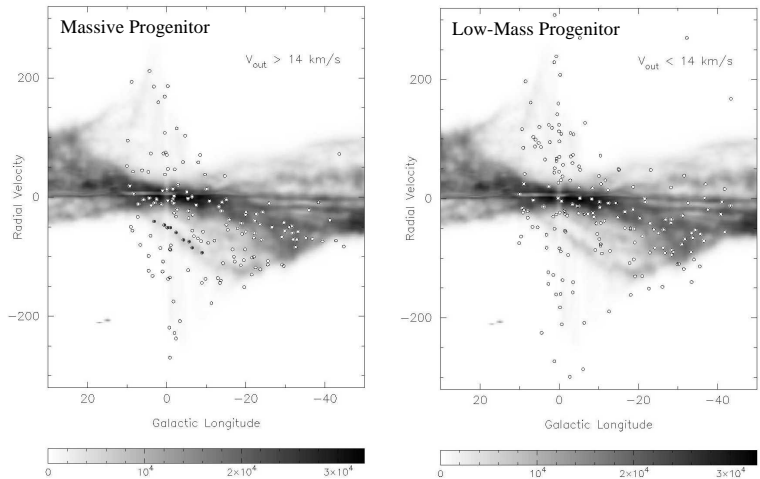}{2.75in}{0}{45}{45}{-190}{-40}
\caption{Plot of Galactic longitude versus velocity for OH/IR stars
near the center of the Galaxy, superimposed on the HI data for this
region.  Those with high outflow velocities have massive progenitors,
and hence must be young; in this group, a population coincident with
the 3~kpc arm feature in the gas, the signature of the gaseous bar, is
highlighted in black.  (Adapted from Sevenster 1999.)}
\end{figure}

It is not yet clear where the stars that currently make up this
triaxial bar formed.  Some have argued that the stars predate the bar,
with little star formation once the bar has formed (e.g.\ Cole \&
Weinberg 2002).  However, there is also strong evidence for continuing
star formation in the bar.  As Figure~5 shows, $l$--$v$ diagrams for
OH/IR stars measured by Sevenster (1999) reveal that those with
massive progenitors (which implies they must be young) pick out the
same 3~kpc arm feature seen in the gas.  Since this feature is a clear
signature of material in an elliptical bar-like structure, it is
apparent that star formation in the bar is an on-going process.

Such continuing star formation suggests that one should view the
puffed-up nature of the stellar bar as arising from a continuing
``heating'' process, in which the stars are progressively scattered
away from the planar gas distribution in which they formed, making
their distribution rounder over time.  Indeed, the lower mass stars in
Figure~5, which are on average older than the more massive stars, show
the somewhat larger scatter in velocity that one would expect from a
heated population.  Another good tracer of intermediate and old
populations is provided by planetary nebulae (PNe), which are
relatively easy to identify from their emission lines, and whose
kinematics can be easily measured from these lines; an analysis of PNe
in the region of the Galactic center (Beaulieu et al.\ 2000) reveals a
distribution in the $l$--$v$ diagram very similar to the older OH/IR
stars in Figure~5, indicating that they have also had extra random
motion injected into their velocities.

A simple model for this extra random component involves adding an
amplitude $\delta$ in quadrature to the gaseous highly-flattened
ellipsoidal distribution, in order to represent the excursions that
the heated stars make from the closed gaseous orbits.  This heuristic
model makes no attempt to explain the details of the scattering
process and the resulting complex distribution function of the stars,
but such an isotropic swelling of the bar is the simplest manner in
which the heating of the stellar component might occur.

\begin{figure}
\plotfiddle{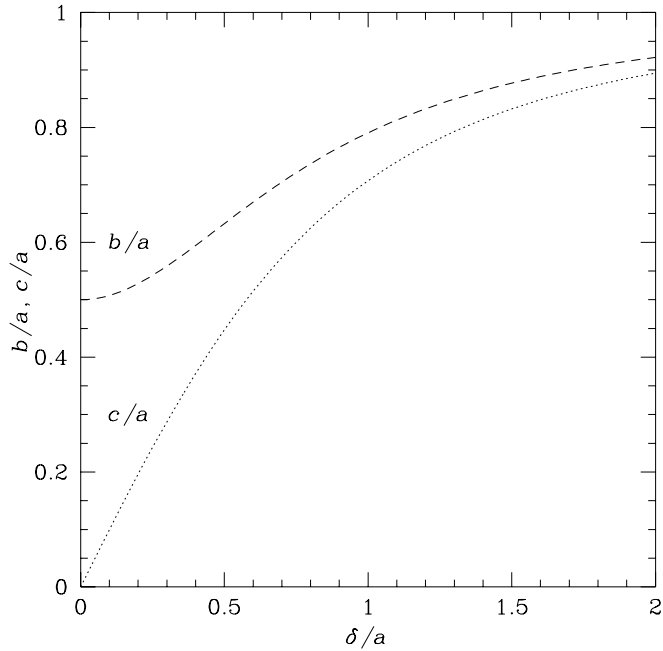}{3.2in}{0}{45}{45}{-140}{-70}
\caption{The effects of a simple model of bar heating, in which a
random amplitude, $\delta$, is added in quadrature to the initial flat
elliptical gas disk, which has an axis ratio $a : b : c \sim 1 : 0.5 :
0$.  Adding $\delta \sim 0.4 a$ reproduces the observed stellar bar
axis ratio of $a : b : c \sim 1 : 0.6 : 0.4$ surprisingly well.}
\end{figure}

Figure~6 shows the effect on the shape of adding progressively larger
random components of amplitude $\delta$ to an initial distribution
shaped like the Milky Way's gaseous component with an aspect ratio $a
: b : c \sim 1 : 0.5 : 0$.  Given the simple-minded nature of this
calculation, it is perhaps surprising that one can simultaneously
reproduce the values of both $b/a \sim 0.6$ and $c/a \sim 0.4$ with a
single value for the amplitude of the random motions, $\delta \sim
0.4a$.  This agreement may just be a fortuitous accident, but perhaps
it sheds some light on the mechanism by which galactic bars become
heated.

\section{The Milky Way as a Barred Galaxy}
We now look beyond the Milky Way to try and place it in the broader
context of spiral galaxies in general.  Again, this is partly just to
satisfy a natural curiosity about our home galaxy.  However, it also
has a broader significance if we are to use the results derived for
the Milky Way to infer anything about galaxies in general: it is
always somewhat risky to draw general inferences from a single object,
but it is downright foolhardy to generalize in this way if we already
know that the Milky Way is in some way unusual.

The first rigorous attempt to classify the Milky Way as if it were an
external galaxy was made by de Vaucouleurs \& Pence (1978), who looked
at a variety of qualitative and quantitative observations of the Milky
Way, and concluded that it should be designated an SABbc(rs) system.
One has to suspect, however, that this large string of characters was
more a reflection of uncertainty than of a detailed understanding of
the Galaxy's structure.  In the case of bar strength, for example, the
paper has very little to say on how the classification was made.  This
reticence is not particularly surprising, as almost nothing was known
about the Galactic bar at that time.  The SAB classification therefore
just reflects the authors' desire to hedge their bets against future
measurements.

\begin{figure}
\plotfiddle{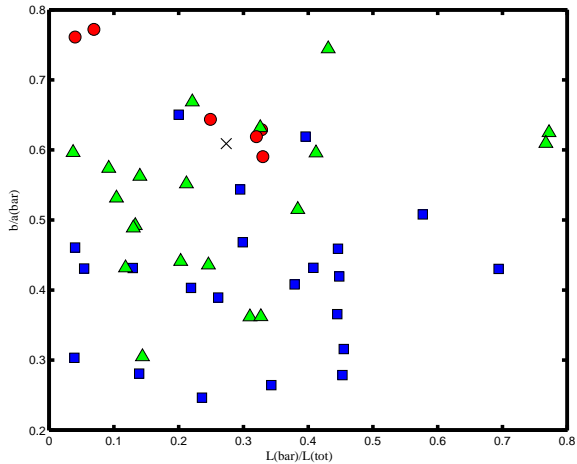}{2.4in}{0}{45}{45}{-120}{0}
\caption{Two-dimensional bar classification, showing the fraction of
total galaxy luminosity of the bar plotted versus bar axis ratio for
the galaxies in the Ohio State Bright Spiral Galaxy Survey data from
the H band. The symbols list the galaxies' conventional bar
classifications of SA (circles) SAB (triangles) and SB (squares).  The
cross is the comparable measurement for the Milky Way.  (Adapted from
Whyte et al.\ 2003.)}
\end{figure}

Twenty-five years on, we have much more data and information,
including the results reviewed in this paper, so we are now in a
position to revisit this question of the Milky Way's classification in
comparison to other galaxies.  We can also now exploit the greater
access to computers and quantitative data in order to move away from
subjective classifications by eye to objective measures of quantities
such as bar strength.  However, even today quantifying bar strength
turns out to be a non-trivial issue, since the term ``strength'' is
not well defined: is a very small, highly elliptical feature at the
center of a galaxy a stronger or weaker bar than a much larger feature
that is closer to axisymmetric?  A variety of measures have been
deployed, ranging from simple observational quantities like
deprojected isophotal flattenings (Martin 1995) to complex
physically-motivated algorithms involving calculating the non-radial
gravitational field of the galaxy (Buta \& Block 2001).

In the current analysis, we recognize the intrinsically
two-dimensional nature of the classification by specifying both an
isophotal ratio for the bar as viewed face on, and the fraction of the
total galaxy luminosity that lies within this isophote.  Figure~7
shows these parameters for the galaxies in the Ohio State Bright
Spiral Galaxy Survey as well as the Milky Way.  It is apparent from
this figure that the conventional SA/SAB/SB classifications of bar
strength are primarily driven by the isophote flattening rather than
the luminosity of the bar, although there is significant scatter in
the quantitative measurements for each of the qualitative classes.
Also, somewhat frustratingly for anyone who would like to classify the
Milky Way definitively into one of these traditional categories, our
galaxy seems to lie in a region of this parameter space where one
finds SA, SAB and SB galaxies.  It might appear that we have not come
very far from de Vaucouleurs and Pence's tentative SAB classification
after all!  However, the uncertainty in the classical classification
of the Milky Way bar now reflects the limitations of these qualitative
schemes rather than any paucity of data on the Milky Way's bar.

\section{The Future}
At some level, the Galactic bar is a solved problem.  As outlined in
this review, there is now a mass of data that provides a coherent
picture of the Milky Way as a barred galaxy.  There are certainly
subtleties that remain to be discovered, but it might reasonably be
argued that bars are not subtle things, and we already know the
salient properties of the Galactic bar to the accuracy that is likely
to be of any interest in studies of galactic structure. 

Although there is some truth in this statement, it also reflects our
ignorance of the fine-scale properties of bars, which arises from the
difficulty of studying these objects in detail at the large distances
of external galaxies.  There is also a good prospect that important
clues to the formation and evolution of bar structures, which are
lacking in the broad-brush picture, can be gleaned from this detailed
information.  The Milky Way therefore offers a unique laboratory for
the detailed study of a bar (albeit one viewed from a rather
disadvantageous direction).

It is already known that bars can have quite complex structures, with
smaller nuclear bars sometimes found within bigger bars at a variety
of orientations [as first noted by de Vaucouleurs (1974)].  These
nested bars are quite common, arising in around a quarter of all
barred galaxies (Laine et al.\ 2002), so it would not be too
surprising to find similar complexity in the arrangement of stars in
the Milky Way's bar.  Indeed, preliminary analysis of the 2MASS star
counts by Alard (2001) has shown that the asymmetry seen in the COBE
data that reveals the presence of a bar (see Section~3) flips in sign
within the central couple of degrees, indicative of a secondary bar at
a radically different orientation.  Our closeness to the Galactic bar
means that there is the prospect of probing much closer to the center
of the Galaxy than is possible in external barred systems, so we might
hope to go down at least two more orders of magnitude in scale to
determine whether this bar-within-a-bar is just the beginning of a
hierarchy of structures nested like Russian dolls.

Huge data sets like the 2MASS star counts are certainly taking the
study of the Milky Way's structure to a new level.  The analysis by
Cole \& Weinberg (2002) showed that the bar can be detected not only
in the asymmetry of the counts, but also more directly by using
infrared carbon stars as standard candles to get a crude measure of
the third dimension.  However, the real revolution will arrive when we
unlock the other three dimensions of phase space by measuring the
velocities of large samples of stars.  There is already some kinematic
data available from studies of line-of-sight velocities and proper
motions of stars in low-extinction windows toward the Galactic center,
and even this limited dynamical information is quite a challenge to
reconcile with detailed models of the bar (H\"afner et al.\ 2000),
although line-of-sight contamination by non-bar stars may well
compromise the analysis.  When the next generation astrometric
satellite GAIA is launched (ESA 2000), we will have measures of both
line-of-sight velocities and proper motions for stars all across the
Galaxy, including those in the bar region, as well as obtaining
parallax distances to these objects to an accuracy of $\sim 10\%$
which will almost eliminate any problems from line-of-sight
contamination.  By assigning ages through position on the
color--magnitude diagram, one will be able to detect the effects of
any secular evolution in the orbit structure [as was done so
successfully to measure evolution in the Galactic disk using data from
GAIA's predecessor Hipparcos (Dehnen \& Binney 1998)].  These data
will push dynamical analysis into a new regime where very
sophisticated modeling will be required to study the details of the
orbit families that are populated in the bar.  The challenges to
modeling and interpreting such huge data sets are far from trivial,
but they will provide a completely new perspective on the formation
and evolution of galactic bars.

The history of astronomy provides a long list of discoveries which
have shown that we are nowhere particularly special in the Universe:
Copernicus moved us out of the center of the Solar System; Shapley
moved the Solar System to a dull suburb of the Milky Way; Hubble
proved that the Milky Way is just one of billions of similar stellar
metropolises.  It should therefore come as no particular surprise to
discover that the Milky Way is like the majority of spiral galaxies in
that it contains a bar.  Although the ordinariness of our situation is
in some ways rather a disappointment, it does have one major up side:
when in the future we want to learn about generic features of the
Universe like bars in disk galaxies, we will not have very far to
look.

\end{document}